\documentstyle[aps,epsf]{revtex}  % DO NOT DELETE THIS LINE 

\global\firstfigfalse  % kludge to bypass foolishness of revtex to place
\global\firsttabfalse  % figures and tables at end of paper

\begin{document}        %  DO NOT DELETE OR CHANGE THIS LINE

\baselineskip 14pt
\title{The CLEO-III RICH Detector and Beam Test Results}

\author{J.C.Wang, M.Artuso, R.Ayad, F.Azfar, E.Dambasuren, 
A.Efimov, S.Kopp, G.Majumder, R.Mountain, S.Schuh, T.Skwarnicki, 
S.Stone, G.Viehhauser}
\address{Syracuse University, Syracuse NY 13244-1130, USA}
\author{S.Anderson, A.Smith, Y.Kubota}
\address{University of Minnesota, Minneapolis MN 55455, USA}
\author{E.Lipeles}
\address{California Institute of Technology, Pasadena CA 91125, USA}
\author{T.Coan, J.Staeck, V.Fadeyev, I.Volobouev, J.Ye}
\address{Southern Methodist University, Dallas TX 75275, USA}

% \author{}   % Use this and the next line only if there is a second
% \address{Another University, etc.}  % address. (Remove the left % marks)
%
\maketitle              % Creates the title area, Do Not Remove

\begin{abstract}        % Do Not Delete this line
We are constructing a Ring Imaging Cherenkov detector (RICH) for 
the CLEO III upgrade for precision charged hadron identification.
The RICH uses plane and sawtooth LiF crystals as radiators,
MWPCs as photon detectors with TEA as the photo-sensitive material,
and low-noise Viking readout electronics.
Results of a beam test of the first two out of total 30 sectors 
are presented.
\
\end{abstract}   	% Do Not Delete this line

\section{Introduction}

The CLEO detector is undergoing a major upgrade to phase III in parallel with a significant
luminosity increase of the CESR electron-position collider\cite{artu96,kopp96}.
These improvements will make it possible to make precision measurements, especially of CP violation
and rare B decays\cite{ston95}.
One of the main goals of the CLEO III upgrade is to have excellent charged
particle identification.

In order to achieve this goal, we have designed and are constructing a Ring Imaging Cherenkov
(RICH) detector, based on the 'proximity focusing' approach\cite{ypsi94}.
It needs no optical focusing elements, and requires that the radiator is relatively thin
compared to the distance of expansion gap between the photon detector and radiator.
The detector can be designed flat and compact to fit in limited space.
Our design is based on the pioneer work done by the Fast-RICH group\cite{arno92,guyo94,moun95}.

When a charged particle passes through a medium, it radiates photons if the velocity
of the particle is faster than light speed in the medium.
The direction of the radiated photons with respect to track is given by
$\cos\theta = 1/(\beta n)$, where $n$ is the index of the medium, and
$\beta = p/E$.
To distinguish between hypotheses for a track at momentum $p$, the formula 
$\Delta \sin^2{\theta} = \Delta m^2 / (p^2\ n^2)$ is applicable.

In a medium of $n=1.5$, a very fast particle  ($\beta \approx 1$) radiates
Cherenkov photons at angle about 840 mrad.
For CLEO III RICH design, our goal is to separate charged pions and kaons
($\rm \pi/K$) at $p$ = 2.65GeV/c, the highest momentum that need be considered at
CLEO for B decays.
The Cherenkov angle difference of $\pi$ and $K$ at this momentum is 14.4 mrad .
In order to obtain very low fake rate with high efficiency, we would like to achieve a 
$4\sigma$ separation between $\pi$ and $K$ at all the momenta relevant for B decays
with the addition of a $1.8\sigma$ dE/dx contribution.
This requires that the RICH provides 4.0 mrad Cherenkov resolution per track.
Thus our benchmark is average 12 photoelectrons per track each with a resolution of $\pm$14 mrad.

\section{Detector Design and construction}

\begin{figure}[ht]	% in second brace, h=here, t=top, b=bottom
\centerline{\epsfxsize=13cm \epsfbox{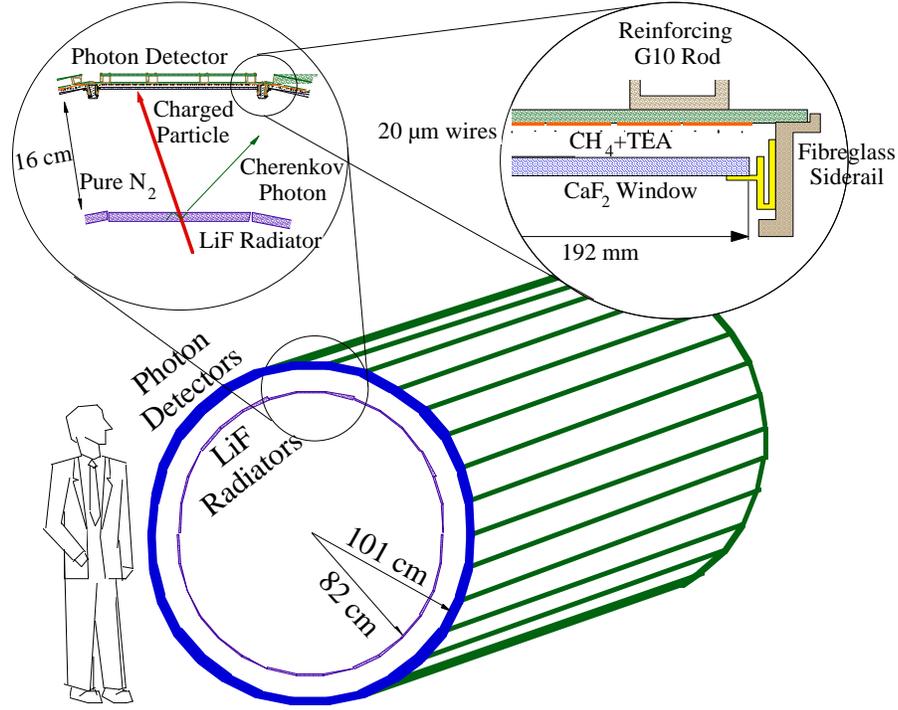}}
\vskip -.2 cm
\caption[]{
\label{richdetector}
\small CLEO III RICH detector. It consists of LiF plane and sawtooth radiators,
       an expansion gap filled with nitrogen, and asymmetric MWPC photon detectors
       with TEA as the photo-sensitive material.}
\end{figure}

The overall structure of the CLEO III RICH detector is cylindrical as shown in 
Fig.~\ref{richdetector}. 
It consists of compact photon detector modules at the outer radius and radiator
crystals at the inner radius, separated by an expansion gap.
The whole detector is divided into 30 sectors in the azimuthal direction.
The RICH resides between the drift chamber and the electromagnetic calorimeter.
The detector occupies a radial space between 80 and 100 cm in radius, is 2.5 m long, and
covers 81\% of the solid angle~\cite{kopp96}.

The choice of the material for the radiator and detector windows
is driven by the choice of Triethylamine (TEA) as the photo-sensitive material.
The $\rm CH_4/TEA$ gas mixture has a finite quantum efficiency in the VUV region
(135-165 nm), which requires the use of fluoride crystals to insure transparency.
The Cherenkov angle resolution is dominated by chromatic dispersion in the emission
of light \cite{efim95}. We choose LiF as radiator for its smaller chromatic dispersion
in VUV region, of all the fluorides. 
The photon detector windows are made of $\rm CaF_2$.

\begin{figure}[ht]
\centerline{\epsfxsize=16cm \epsfbox[0 0 580 120]{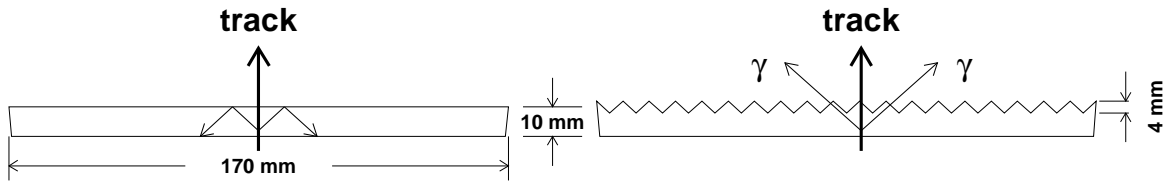}}
\vskip -.2 cm
\caption[]{
\label{radiator}
\small Azimuthal view of plane and sawtooth radiators. For normal incident tracks,
      the Cherenkov photons are trapped in the plane radiator, whereas they can 
      escape the sawtooth radiator.}
\end{figure}

The LiF radiators will be mounted on an inner carbon fiber cylinder as a$\rm 14 \times 30$ array,
each radiator is $\rm 170 \times 170 mm^2$ and 10 mm thick.
The normal choice of radiator shape is a flat plate.
However, as the total reflection angle of 150 nm photon in LiF 
($n \approx 1.5$ at 150 nm) is about $42^\circ$ and the Cherenkov angle is close to
$48^\circ$, the photons radiated by track at normal incidence  will be trapped as shown
in Fig.~\ref{radiator}.
Instead of tilting the plates to allow the light to get out near the center of the detector,
we developed radiators with a 'sawtooth' pattern
on its outer surface, which allows the Cherenkov photons to escape the radiator\cite{efim95}.
We will use such sawtooth radiators for the center four of the 14 rings.

The expansion gap (16 cm) is filled with high purity nitrogen gas.
The volume is well sealed to minimize the contamination of $\rm O_2$ and
$\rm H_2O$, which absorb VUV photons.

The photon detector is multi-wire proportional chamber with cathode pad readout.
It is filled with $\rm CH_4$ gas bubbled through liquid TEA at $\rm 15^\circ C$ (7\%).
The entrance windows are 2 mm thick $\rm CaF_2$,
coated with 100 $\rm \mu m$ wide silver traces to act as an electrode.
Photons with a wavelength between 135 nm and 165 nm can generate single
photoelectrons in the $\rm CH_4/TEA$ mixture.
The photoelectron drifts toward 20 $\rm\mu m$ diameter Au-W anode wires,
where avalanche multiplication takes place.
Charge is induced on an array of $\rm 8.0\times7.5\ mm^2$ cathode pads,
allowing a precise reconstruction of the position of the Cherenkov photon at the
detector plane.
The chamber is asymmetric, with the wire to pad distance of 1 mm and the wire to
window distance of 3.5 mm, to improve the wire cathode pad coupling.

\section{Electronics}

\begin{figure}[ht]
\centerline{\epsfxsize=11cm \epsfbox{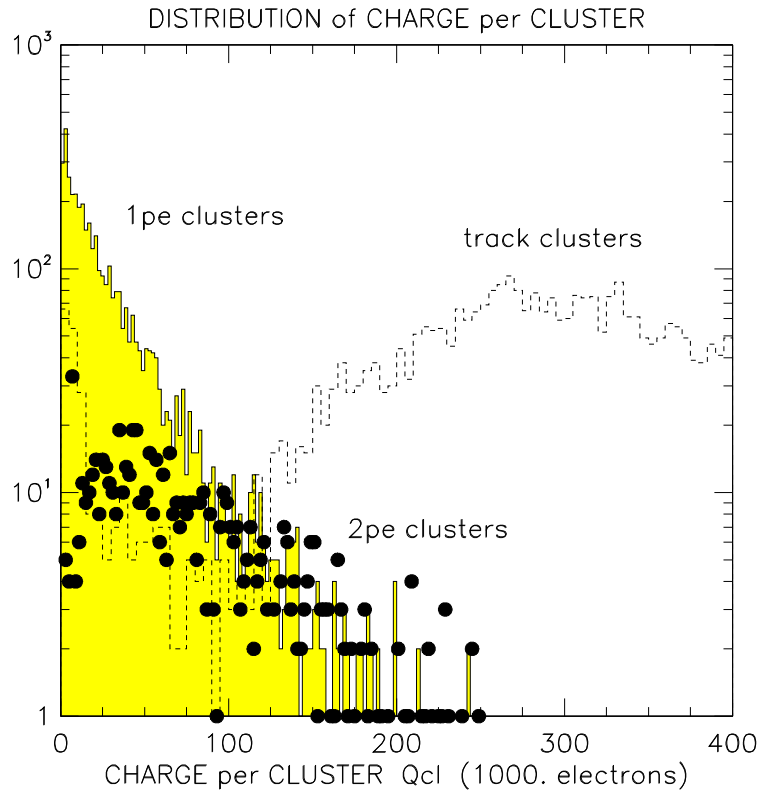}}
\vskip -.2 cm
\caption[]{
\label{charge}
\small Distribution of charge per reconstructed cluster. The distribution of single
photoelectrons is exponential.}
\end{figure}

The choice of the readout electronics for CLEO III is governed by several
Considerations \cite{artu96}.
First, the charge induced by a single photoelectron avalanche at moderate gain
follows an exponential distribution as shown in Fig.~\ref{charge}.
As the most likely charge is zero, it is necessary to have a very low noise
system so that high efficiency is achieved.
Furthermore, we want to reconstruct charged tracks accurately.
We expect the pulse height from the charged track to be at least a factor 20 higher
than the mean gain for a single photoelectron.
In addition, high resolution analog readout allows a more effective suppression of cluster overlaps. 
Lastly, in the final system, we will have 230,400 channels to be readout.
An effective zero suppression algorithm is necessary to have a manageable data size.

We use the VA\_RICH chip as the front end processor. It is a custom-designed
64-channel VLSI chip based on the VA series developed for Si microchip readout \cite{nyga91}.
It features low noise and large dynamic range.
With a 2 pF input capacitor, its equivalent noise charge is about 150 electrons.
Linearity is excellent up to $\rm \pm 4.5 \times 10^5 e^-$ input.
Two chips are wire bonded to one hybrid circuit.
Sixty hybrids are mounted on each photon detector.

The signal travels over a 6 m long cable from the detector to a VME databoard.
The databoard includes receivers, 12-bit differential ADCs, bias circuitry for the VA\_RICH
chips, a circuit generating the timing sequence needed by the VA\_RICH chip, and VME interface.
In the final version, the databoards will include a DSP based algorithm for common mode
subtraction that will make the system less vulnerable to coherent noise sources.
For the beam test, 8 prototype databoards in one crate are used.
These databoards were not equipped with coherent noise subtraction,
and all the channels were readout to monitor and correct for coherent noise fluctuations.

\section{Beam Test setup}

\begin{figure}[ht]
\centerline{\epsfxsize=14cm \epsfbox{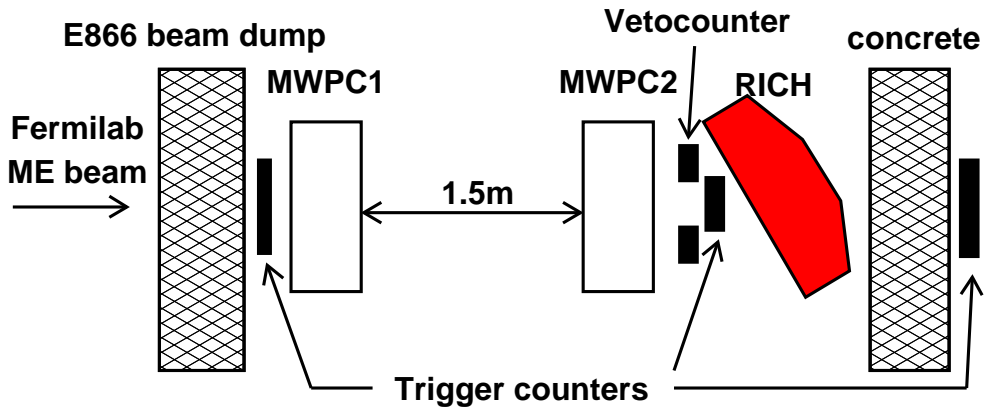}}
\vskip -.2 cm
\caption[]{
\label{beamtest}
\small Beam test setup. The two photon detectors and radiators are mounted in the RICH box to
       simulate final configuration. MWPCs are used as reference tracking system. 
       Scintillator counters
form trigger system.}
\end{figure}

In order to test our design of the RICH detector, a comprehensive beam test was performed.
%%In this test, we simulated 1/15 of the final CLEO III RICH.
The beam test  setup is shown in Fig.~\ref{beamtest}.

The first two photon detectors built for the final CLEO III RICH  were mounted in an 
aluminum box.
In addition,  one plane and two sawtooth LiF radiators were mounted in the box
at the same distance from the photon detectors as designed for CLEO III RICH.
The box was sealed and flushed with pure nitrogen.
A schematic of the box can be found in Ref.~\cite{vieh98}.

The beam test was performed in a muon halo beam in the Meson East area
of FermiLab.
Apart from the RICH itself, the system consisted of 
trigger scintillator counters, and a tracking system.
We used two sets of MWPCs to measure the track position and angle. They provided
0.7 mm spatial resolution per station and a track angle resolution of 1 mrad.
To simulate different track incidence angles,
the RICH detector box was rotated at various polar and azimuthal angles.

\section{Beam test electronics performance}

During the three week beam test, the photon detectors were run at an average gain of
$4 \times 10^4$.
The total electronic noise was about 1000 electrons.
After coherent noise subtraction, the residual incoherent noise was 400 electrons,
providing an average signal-to-noise ratio for single photon signal of 100:1.

\begin{figure}[ht]
\centerline{\epsfxsize=7cm \epsfbox{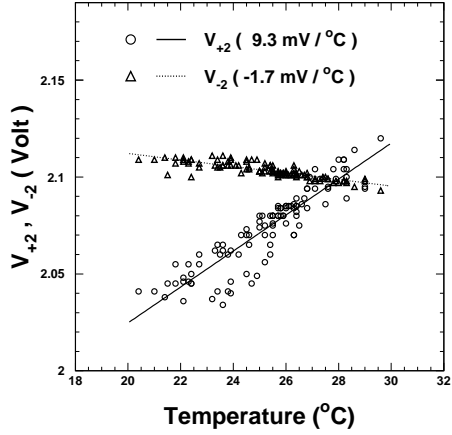}}
\vskip -.2 cm
\caption[]{
\label{volt_t}
\small Temperature sensitivity of $\pm 2$ V power on databoard.}
\end{figure}

An important effect that we discovered with the beam test was the temperature sensitivity
of the analog +2V and --2V, necessary to bias the VA\_RICH.
Fig.~\ref{volt_t} shows this effect.
This, in turn, affected the bias configuration of the chip and its pedestal.
The $\pm 2$ voltage regulator design has been changed to eliminate this problem.

\section{Results}

\begin{figure}[ht]
\centerline{\epsfxsize=8cm \epsfbox{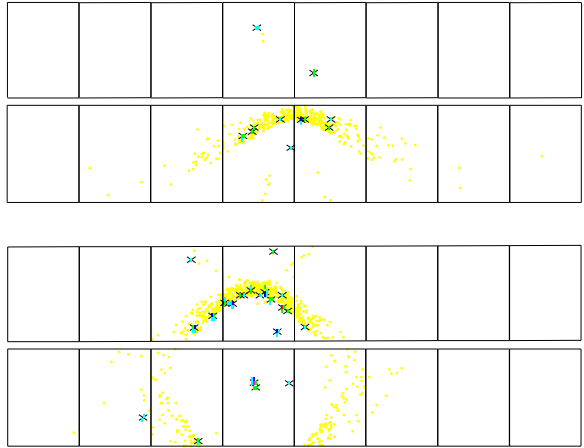}}
\vskip -.2 cm
\caption[]{
\label{event}
\small Event displays. The 'x' indicate reconstructed Cherenkov photons 
       The dots are expectations of a Monte Carlo simulation.
       For plane radiator with $30^\circ$ incident track (upper), the one-arc image is visible.
       For sawtooth radiator with normal incident track (lower), two arc images are visible with
       the lower one truncated due to limited acceptance.}
\end{figure}

%%Data has been taken at a variety of track angles by rotating the RICH box.
Fig.~\ref{event} shows the displays of two single events from the plane radiator
at $30^\circ$ track incidence and from a sawtooth radiator at normal incidence.
For the plane radiator, the image is a single arc as shown.
While for the sawtooth radiator, two arcs on opposition sides of the charged track are visible,
with the lower one truncated due to acceptance.
The acceptance for images from plane radiators is about 85\% of the maximum possible acceptance. 
With the necessary chamber mounts and window frames, 85\% is the maximum possible acceptance. 
The acceptance for sawtooth images was about 50\% in this setup, and should
be about 85\% in the final system.

Channels recording pulse height above a threshold of $5\sigma$ noise are selected 
for further analysis.
The first step in the analysis is clustering.
The center of the cluster is treated as the location of a photoelectron.
About 2.2 pads per cluster are found, with 1.1 photoelectrons per cluster, due to some
unresolved photon overlaps.
For each photoelectron, the trajectory is optically traced back through all media,
with the assumption that it originates from the mid-point of radiator.
From this propagation path, the Cherenkov angle is calculated.

\begin{figure}[ht]
\centerline{\epsfxsize=13cm \epsfbox{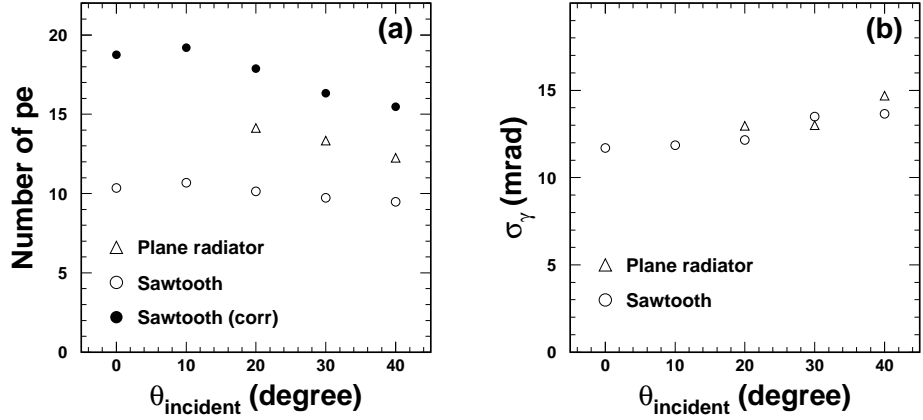}}
\vskip -.2 cm
\caption[]{
\label{photon}
\small Average number of photoelectrons per track (a) and Cherenkov angle resolution per
       photoelectron (b).
       For sawtooth radiator, number of photoelectrons is is estimated for the
       final RICH system shown in dots (The plane radiator already has full acceptance).}
\end{figure}

We extract the number of photoelectrons per track by fitting the single photon spectrum
taking into account the background.
We invoke a $\pm 3 \sigma$ cut so as not to include non-Gaussian tails.
The numbers of photoelectron per track at different incident angle are shown in 
Fig.~\ref{photon} (a).
For plane radiator, the average number of photoelectrons per track is between 12 
and 15, varying with incident angle.
For the sawtooth radiator the number is lower due to limited acceptance.
An extrapolation made to estimate the performance of the final system predicts
17-20 photoelectrons.
Fig.~\ref{photon} (b) shows the measured Cherenkov angle resolution per photon $\sigma_{\gamma}$.
The resolution is between 11 and 15 mrad.

\begin{figure}[ht]
\centerline{\epsfxsize=13cm \epsfbox{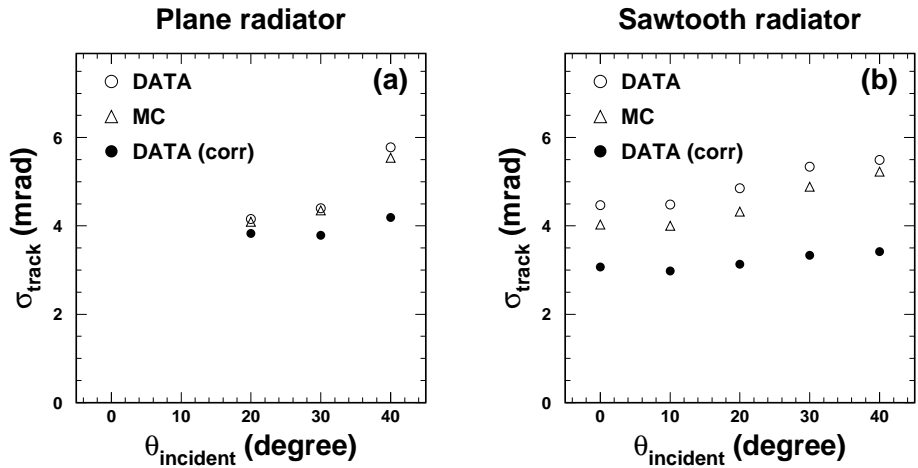}}
\vskip -.2 cm
\caption[]{
\label{track}
\small Per track Cherenkov angle resolution for (a) plane and (b) sawtooth radiators as
       a function of track incident angles. Open circles are measured beam test data, the
       triangles show Monte Carlo simulated data. 
       Also shown as filled circles are resolutions derived from beam test results extrapolated
       to the acceptance in the final system that exclude the tracking error contribution. 
       The final resolution in CLEO III is expected to be between open circles and filled circles.}
\end{figure}

The parameter that shows the particle identification power of this system is the
Cherenkov angle per track $\sigma_{track}$.
The resolutions are summarized in Fig.~\ref{track} for both plane and sawtooth radiators.
The measured $\sigma_{track}$ increases with track angle\cite{efim95}.
A Monte Carlo study is performed to simulate the resolution.
The Monte Carlo represents the plane radiator data well.
For the sawtooth radiator, the data show a worse resolution than the simulation.
There are several sources in the simulation that could account for this discrepancy, such as the
tracking error, beam background, and imperfect knowledge of the sawtooth shape.

The Monte Carlo study also shows that the contribution to $\sigma_{track}$ from the
tracking errors is significant,
especially for large incident track angles.
This error comes in due to the photon emission point error.
For the plane radiator at $40^\circ$, the contribution is about 4 mrad, while
the total error is about 5.8 mrad.
In the final CLEO III system, the tracking system is expected to be substantially better
than the system used in the beam test.
In order to have approximate projection of the performance of the CLEO III RICH, we extrapolate
the resolution from the beam test results, with the assumption of no tracking
error and the acceptance in final system. This resolution is shown as the filled circles in Fig.~\ref{track}.
%%The expected CLEO III RICH performance will be worse than this extrapolation.

\section{Summary and outlook}
The beam test results indicate that the specifications of CLEO III RICH design
are fulfilled.
In particular, the Cherenkov angle resolution per track reaches 4 mrad,
which will provide $\rm 4\sigma\ \pi/K$ separation in CLEO III \cite{Raym}.
The RICH detector is on the final stage of construction.
We expect the completion  of the project in Summer 1999.

\section{Acknowledgments}

We would like to thank FermiLab for providing us with necessary infrastructure and
dedicated beam time. We give particular thanks to Chuck Brown and other colleagues from
E866 for their hospitality.

\end{document}